\documentclass[usenatbib]{basi}
\usepackage[T1]{fontenc}
\usepackage[british]{babel}
\usepackage[varg]{txfonts}
\usepackage{rotating}
\usepackage{dcolumn}

\def\ppt{t^{\prime \prime}}

\def\lp{{\ell}_{\rm Pl}}

 \newcommand{\rcr}{\rho_{\mathrm{crit}}}

\newcommand{\p}{\partial}

\newcommand{\f}{\frac}

\def\rmax{\rho_{\mathrm{max}}}

\def\rcr{\rmax}

\def\f{\frac}

\def\d{\textrm{d}}

\def\ul{\underline}

\newcommand{\be}{\nopagebreak[3]\begin{equation}}
\newcommand{\ee}{\end{equation}}
\newcommand{\ba}{\nopagebreak[3]\begin{eqnarray}}
\newcommand{\ea}{\end{eqnarray}}

\newcommand{\bmult}{\nopagebreak[3]\begin{multline}}
\newcommand{\emult}{\end{multline}}

\def\d{{\rm d}}

\def\ppt{t^{\prime \prime}}

\def\lp{{\ell}_{\rm Pl}}

\def\f{\frac}

\def\d{\textrm{d}}

\def\ul{\underline}

\def\rmax{\rho_{\rm max}}

\def\sina{\sin{(\bar{\mu}_1 c_1)}}

\def\sinb{\sin{(\bar{\mu}_2 c_2)}}

\def\sinc{\sin{(\bar{\mu}_3 c_3)}}

\def\cosa{\cos{(\bar{\mu}_1 c_1)}}
\def\cosb{\cos{(\bar{\mu}_2 c_2)}}

\def\lp{l_{\rm Pl}}

\def\d{{\rm d}}

\begin{document}
\title[Loop quantum cosmology and the fate of cosmological singularities]{Loop quantum cosmology and the fate of cosmological singularities}
\author[Parampreet Singh]%
       {Parampreet Singh\thanks{email: \texttt{psingh@phys.lsu.edu}; This invited review article is based on the Professor M. K. Vainu Bappu Memorial Gold Medal 2010 Award Lecture.}\\
       Department of Physics and Astronomy, Louisiana State University\\
       Baton Rouge, Louisiana 70803, USA}

\pubyear{2014}
\volume{42}
\pagerange{121--146}
\date{}
\maketitle
\label{firstpage}

\begin{abstract}
Singularities in  general relativity such as the big bang and big crunch, and exotic singularities such as the big rip are the boundaries of the classical spacetimes. These events are marked by a divergence in the curvature invariants and 
the breakdown of the geodesic evolution. Recent progress on implementing techniques of loop quantum gravity to cosmological models reveals 
that such singularities may be generically resolved because of the quantum gravitational effects. Due to the quantum geometry, which 
replaces the classical differential geometry at the Planck scale, the big bang is replaced by a big bounce without any assumptions on the 
matter content or any fine tuning. In this manuscript, we discuss some of the main features of this approach and the results on the generic 
resolution of singularities for the isotropic as well as anisotropic models. Using effective spacetime description of the 
quantum theory, we show the way quantum gravitational effects lead to the universal bounds on the energy density, the Hubble rate and the 
anisotropic shear. We discuss the geodesic completeness in the effective spacetime and the resolution of all of the strong singularities. It turns out that despite the bounds on energy density and the Hubble rate, there can be divergences in the curvature invariants. However such events are geodesically extendible, 
with  tidal forces not strong enough to cause inevitable destruction of the in-falling objects.

\end{abstract}


\section{Introduction}

Einstein's theory of general relativity is extremely successful in describing the evolution of our universe from the early stages to the 
very large scales. However, it suffers from the problem of classical singularities which are the generic features of 
spacetimes in general relativity. In the cosmological context, a  spatially flat Friedmann-Lema\^itre-Robertson-Walker (FLRW) spacetime 
filled with matter with an equation of state such as of dust, radiation, a stiff fluid or even an inflaton, evolves from a big bang singularity in 
the past where the scale factor vanishes. If the universe is contracting, then for these types of matter,  evolution ends in a big crunch 
singularity in a finite future. The physics of these singularities is very rich. First, let us note that in general not all singularities occur at a vanishing scale factor. A singularity may occur even at an 
infinite value of the scale factor, such as in the big rip, or at a finite value of the scale factor such as in the sudden and big freeze 
singularities. Further, singularities come in different strengths and they can be strong or weak 
\citep{ellis-schmidt,clarke-krolak,tipler,krolak}. Strong singularities are identified by the tidal forces which are infinite, causing 
inevitable destruction 
of any in-falling object with arbitrary characteristics. Weak singularities on the other hand can have divergence in the spacetime 
curvature, but they are not strong enough to destroy arbitrary in-falling detectors. Big bang and big rip are examples of the strong 
singularities where geodesic evolution breaks down and hence these are the boundaries of classical spacetime. Examples of weak 
singularities include sudden singularities, beyond which geodesics can be extended. In the presence of anisotropies, such as in the 
Bianchi-I model, singularities come in different `shapes.' \citep{dorosh,thorne,ellis,ellis1,jacobs}. Unlike the point like big 
bang/big crunch singularity in the isotropic cosmological models, in the presence of anisotropies the singularities can also be a cigar, 
barrel, and a pancake like depending on the behavior of directional scale factors.   If the spatial curvature is present in 
addition, the approach to classical singularities oscillatory and leads to the mixmaster dynamics 
\citep{berger,garfinkle}.

The existence of singularities, such as the big bang, shows that classical general relativity reaches its limits of validity when the 
spacetime curvature becomes extremely large, and in such a case gravitational physics should be described by a 
quantum theory of gravity. One of the main attempts to quantize gravity is loop quantum gravity, which is a non-perturbative canonical 
quantization of gravity based on Ashtekar variables \citep{report,rovelli,thiemann,pullin}.  Though a full theory of loop quantum gravity is 
not yet available, it has reached a high mathematical precision and various important results have been obtained. A key prediction of loop 
quantum gravity is that classical differential geometry of general relativity is replaced by a quantum geometry at the Planck scales. For 
spacetimes with symmetries, such as the cosmological and black hole spacetimes, techniques of loop quantum gravity have 
been used to perform a rigorous  quantization and detailed physical implications have been studied.
In this manuscript we focus on the applications of these techniques to cosmological models in loop quantum cosmology, where isotropic and 
anisotropic models have been widely investigated, and recently these 
techniques have been used to study the effects of quantum gravity in the inhomogeneous situations, including Gowdy models \citep{gowdy1,gowdy2,gowdy3} and quantum fluctuations in loop quantized spacetime \citep{aan1,aan2}. Loop quantization of black hole spacetimes, a discussion of which falls beyond the scope of this manuscript, uses similar techniques as in loop quantum cosmology, and leads to similar results on singularity resolution  \citep{ab_bh,gambini1,gambini2}.

Loop quantum cosmology carries forward the quantum cosmology program which started from the canonical quantization of cosmological spacetimes 
in the metric variables. In the metric based formulation of quantum cosmology, the Wheeler-DeWitt quantum cosmology, 
 resolution of singularities remains problematic. The spacetime in Wheeler-DeWitt quantization is 
a continuum as in the classical general relativity and strong singularities are present in general unless one chooses very special 
boundary conditions. In a sharp contrast, the underlying geometry in loop quantum cosmology is a discrete quantum  geometry inherited from 
 loop quantum gravity. Instead of the 
differential equation which governs the evolution in the Wheeler-DeWitt theory, in loop quantum cosmology, evolution operator is a discrete 
quantum operator \citep{mb,abl}. It is only at the large scales compared to the Planck scale, that the discrete quantum geometry is approximated by the 
classical differential geometry, and there is an agreement between the physical implications of loop quantum cosmology and 
the Wheeler-DeWitt theory. However, in the Planck regime there are significant and striking differences between Wheeler-DeWitt theory and 
loop quantum cosmology. A sharply peaked state on a classical trajectory when evolved towards the big bang in the Wheeler-DeWitt theory, 
follows the classical trajectory all the way to the classical singularity. In loop quantum cosmology, such a state bounces when energy 
density reaches a maximum value $\rcr \approx 0.41 \rho_{\rm{Pl}}$ \citep{aps1,aps2,aps3}. The occurrence of bounce which, first observed in the loop 
quantized spatially flat FLRW model with a massless scalar field, occurs without any violation of the energy conditions or fine tuning. The physics  of the bounce  has been found to be robust using an exactly 
solvable model \citep{slqc}. The quantum probability for the bounce turns out to be unity \citep{craig-singh2}.\footnote{In contrast, the quantum probability for bounce in Wheeler-DeWitt theory for this model is zero \citep{craig-singh1}.} Following the method for the spatially flat model, loop quantization has been carried out for different matter models and in the presence of spatial curvature in homogeneous spacetimes   
\citep{apsv,warsaw,bp,ap,ck-closed,rad,aps4,cyclic}. The existence of bounce is a generic result in all these investigations. The states remain sharply 
peaked throughout the non-singular evolution and quantum fluctuations are tightly constrained across the bounce \citep{cs1,triangle,cm1}. 
The quantum resolution of big bang singularity is not confined only to the isotropic models. The quantum evolution operator in anisotropic 
models, such as the Bianchi-I, Bianchi-II and Bianchi-IX models, has been shown to be non-singular 
\citep{chiou,madrid_bianchi1,awe2,awe3,we,pswe}. In case of the infinite degrees of freedom, Gowdy models have been recently studied. Though these models are not yet fully loop quantized, important insights on the 
nature of bounce have been obtained \citep{gowdy1,gowdy2,gowdy3}. 

An important feature of loop quantum cosmology is the effective spacetime description of the underlying quantum evolution. For the states which lead 
to a classical macroscopic universe at late times, this description can be obtained from an effective Hamiltonian \citep{jw,vt}. Rigorous 
numerical simulations have confirmed the validity of the effective dynamics \citep{aps3,apsv,bp,ap,dgs2}, which provides an excellent 
approximation to the full loop quantum dynamics for the sharply peaked states at all the scales. It has been shown that only when the 
states have very large quantum fluctuations at late times, i.e. they do not lead to macroscopic universes described by general relativity, 
that the effective dynamics has departures from the quantum dynamics near bounce and the subsequent evolution \citep{dgs2,squeezed}. In such a case, the effective dynamics overestimates 
the density at the bounce, but still captures the qualitative aspects extremely well. The effective dynamics approach has been extensively 
used to study physics at the Planck scale and the very early universe in loop quantum cosmology (see \citet{asrev} for various 
applications of the effective theory). Some of the applications include 
non-singular inflationary attractors in isotropic \citep{svv,as1,multi} and anisotropic models \citep{gs2}, and the probability of 
inflation to occur \citep{as1,as2,ck,corichisloan}, transitions between non-singular geometrical  structures  in the Bianchi-I model \citep{gs3}, singularity resolution in string 
cosmology inspired  \citep{pbb,csv}, and multiverse scenarios \citep{multiverse1,gs4}, and various studies on cosmological perturbations (see 
\citet{barrau} for a review). 

An interesting application of effective dynamics is to study the fate of singularities in general for different matter models. The idea is
 based on analyzing the generic properties of curvature invariants, geometric scalars such expansion and shear scalars, and the geodesic evolution in the effective spacetime. This approach leads to some striking results. Assuming the validity of the effective dynamics, one finds that for generic matter with arbitrary equation of state, the effective spacetime in the spatially flat model is geodesically complete, and there are no strong curvature singularities \citep{ps09}.   Loop quantum cosmology resolves all the singularities such as big bang, big rip and big freeze \citep{ps09,sst,gumju,naskar}, but interestingly ignores the sudden singularities \citep{portsmouth,ps09}. It is interesting to note that quantum geometric effects are able to differentiate between genuine and the harmless singularities, and only resolve the former types. These results hold for the case when spatial curvature is present, where resolution of strong singularities and non-resolution of weak singularities has been confirmed via a 
phenomenological analysis \citep{psfv}. Generalization of this analysis has been performed in the presence of anisotropies where it has been shown that all strong singularities which occur in classical general relativity in the Bianchi-I spacetime are avoided, and the geodesic evolution does not break down in general \citep{ps11}.  These results are tied to the universal bounds on the 
 energy density, expansion scalar (or the mean Hubble rate), and shear scalar in the effective spacetime of loop quantum cosmology for the isotropic and Bianchi-I model \citep{ps11,cs3}. Recently, similar bounds on geometric scalars have been obtained for the Bianchi-II \citep{gs1}, Bianchi-IX \citep{pswe} and Kantowski-Sachs spacetime \citep{js1}, which indicates that these conclusions can be generalized to other anisotropic models and black hole spacetimes. 

In this manuscript, we give an overview of the main result of the bounce, and the generic resolution of singularities using effective spacetime description in loop quantum cosmology. To make this manuscript self contained with basic techniques, in Sec. 2 we summarize concepts 
in the classical theory which are needed for discussion in the later sections. Loop quantum cosmology is a canonical quantization involving a $3+1$ decomposition of the spacetime. 
This decomposition and the Hamiltonian framework is summarized in the first part of Sec. 2. We express the Einstein-Hilbert action in terms of the quantities defined with respect to the three dimensional spatial slices, and obtain constraints both in the metric variables and Ashtekar variables in the classical theory. We then implement these techniques to obtain a Hamiltonian formulation of the spatially flat FLRW spacetime and show the way classical field equations, such as the Friedmann equation, can be obtained from the Hamiltonian constraint  and the Hamilton's equation. In Sec. 2.3 and 2.4, we discuss geodesic evolution, strength and different types of singularities. Though these are discussed in the isotropic setting, the same classifications apply to the anisotropic spacetime. In Sec. 3, we provide a short summary of the loop quantization of the spatially flat homogeneous and isotropic model in loop quantum cosmology. We discuss the way quantum geometric effects lead to a discrete quantum evolution 
equation which avoids the singularity, and leads to the quantum bounce of the universe at the Planck scale. Sec. 4 deals with the effective spacetime description and application of this method to understand bounds on Hubble rate, curvature scalars, geodesic completeness and the absence of strong curvature singularities. A discussion of these results to an anisotropic spacetime, the Bianchi-I model, is provided in Sec. 5. We conclude with a brief discussion of the main results in Sec. 6. Due to the space constraints, it is is not possible to 
 cover all the details and developments in this manuscript, especially various conceptual issues and the progress in numerical methods which have provided valuable insights and served as robustness tests on singularity resolution. An interested reader may refer to reviews, such as \citet{report} and \citet{asrev} for details on loop quantum gravity and loop quantum cosmology respectively. The numerical techniques used in loop quantum cosmology are reviewed in detail in \citet{ps12} and \citet{khanna}.

\section{Classical theory: Hamiltonian framework and and the properties of singularities}
In this section, we overview some aspects of the classical theory which are useful to understand the underlying procedure of quantization 
of cosmological spacetimes in loop quantum cosmology, and the properties of singularities. We start with the $3+1$ decomposition of the 
classical spacetime and discuss the way Hamiltonian and diffeomorphism constraints arise in the metric variables and the Ashtekar 
variables. 
We then specialize to the case of the spatially flat homogeneous and isotropic model, where due to the symmetries the diffeomorphism constraint is satisfied, and the only 
non-trivial constraint is the Hamiltonian constraint. Using Hamilton's equations, we demonstrate the way classical Friedmann equation can be 
 derived in this setting. In the last part of this section, we discuss the properties of geodesics and different types of singularities in 
the classical theory for the $k=0$ FLRW spacetime.

\subsection{The 3+1 decomposition and constraints}
Loop quantum gravity is a canonical quantization of gravity based on a  $3+1$ decomposition of a spacetime manifold. Let us consider the  topology of the spacetime with metric $g_{ab}$ as  $\Sigma \times R$. If 
there exists a global time $t$, then the spacetime 
can be foliated into constant time hypersurfaces $\Sigma$, each with a spatial metric 
\be
q_{ab} = g_{ab} + n_a n_b ~,
\ee
where $n_a$ is a unit normal to the hypersurface $\Sigma$. Using it, a time-like vector field $t^\mu$ can be decomposed into normal and tangential components to 
the spatial slices $\Sigma$ as $t^a = N n^a + N^a$, where $N$ is the lapse and $N^a$ denotes the shift vector. 
In the Hamiltonian formulation of general relativity in the metric variables, $q_{ab}$ acts as the configuration variable. Its conjugate variable is $\pi^{ab} = \sqrt{q} (K^{ab} - K q^{ab})$, 
where $q$ denotes the determinant of the spatial metric, and $K_{ab}$ is the extrinsic curvature of the spacetime 
defined as $K_{ab} = - q_a^c \nabla_c n_a$. 

Using the $3+1$ decomposition, the Einstein-Hilbert action for general relativity can be expressed as 
\ba
S &=& \nonumber  \int \d^4 x \sqrt{-g} R \\ &=& \int \d^4 x \left( \pi^{ab} \dot q_{ab} + N \sqrt{q} \left( {}^{(3)}R - q^{-1} (\pi^{cd} \pi_{cd} + \f{1}{2} (\pi^i_i)^2) \right)  + 2 N_a \sqrt{q} D_b(\sqrt{q} \pi^{ab}) \right) ~,
\ea
where $D_a$ denotes the 3-dimensional covariant derivative.  Note that the Lagrangian does not contain any time derivatives of the lapse $N$ 
and the shift vector $N_a$. Therefore, these act as Lagrange multipliers, variations with respect to which yield us 
the constraints:
\be
{\cal C}_H = -{}^{(3)}R + q^{-1} (\pi^{ab} \pi_{ab} - \f{1}{2} (\pi^i_i)^2) = 0
\ee
and 
\be
{\cal C}_D^a = D_b (\sqrt{q} \pi^{ab}) = 0 ~.
\ee
Here $C_H$ is the Hamiltonian constraint and $C_D^a$ is the spatial diffeomorphism constraint. 
The total Hamiltonian is the sum of these constraints: ${\cal H} = \int \d^3 x (N {\cal C}_H + N_a {\cal C}^a)$. The form of the Hamiltonian 
in the metric variables makes it very difficult to obtain physical solutions in the quantum theory. Nevertheless, for spacetimes with 
symmetries, such as homogeneous cosmological spacetimes, one can quantize the Hamiltonian and obtain physical states. This metric based 
approach to quantize cosmological spacetimes is known as the Wheeler-DeWitt quantum cosmology, some aspects of which will be discussed in the 
next section. 

The Hamiltonian constraint becomes simpler to handle in the Ashtekar variables \citep{abhay}: the connection $A^i_a$ and the 
(densitized)\footnote{Given a triad $e^a_i$, the densitized triad is obtained by $E^a_i = \sqrt{|q|} e^a_i$.} triad $E^a_i$, where $a$ and 
$i$ are the spatial and internal indices respectively, taking vales 1,2,3. The spatial metric on $\Sigma$ can be constructed using triads 
as,
\be
E^a_i  E^b_j \delta^{ij} =  |q| q^{ab} ~.
\ee
The connection is related to the time derivative of the metric, and can be written in terms of the extrinsic curvature as
\be
A^i_a = \Gamma^i_a + \gamma K^i_a ~,
\ee
where $\gamma \approx 0.2375$ is the Barbero-Immirzi parameter whose value is fixed in loop quantum gravity using black hole 
thermodynamics, $K_a^i = K_{ab}  E^{ai}/\sqrt{|q|}$, and $\Gamma^i_a$ is the spin connection:
\be
\Gamma^i_a = -\f{1}{2} \epsilon^{i j k} e_j^b(\partial_a e_b^k - \partial_b e_a^k + \delta_{m n} e_k^l e_a^m \partial_l e_b^n)
\ee
where $e^a_i = E^a_i/\sqrt{|q|}$, and $e^a_i e_a^j = \delta^j_i$. 

Rewriting the Einstein-Hilbert action in terms of the Ashtekar variables, one obtains the following constraints:
\be
D_a E^a_i = 0, ~~~~~~ E^a_i F_{ab}^i = 0  
\ee
and
\be\label{hamclassical}
 \f{1}{\sqrt{|q|}}  E^a_j E^b_k (\epsilon_{i}^{j k} F_{ab}^i - (1 + \gamma^2) (K^j_a K_b^k - K^j_b K_a^k)) = 0~.
\ee
The first constraint is the Gauss constraint, the second is the momentum constraint and the third constraint is the Hamiltonian constraint. Here the latter two are obtained by imposing the Gauss constraint.  The Gauss and the momentum 
constraints can be combined to give the diffeomorphism constraint. Once we have the Hamiltonian as the linear combination of the 
constraints, we can obtain dynamical equations from the Hamilton's equations which along with the constraint equations yield us an 
equivalent formulation of the Einstein field equations. In the quantum theory, the physical states are those which are annihilated by the operator corresponding to 
quantum Hamiltonian. These states constitute  the physical Hilbert space of the quantized gravitational spacetime.  

We conclude this part by noting a conceptual difficulty in the Hamiltonian framework. Since the Hamiltonian is 
a linear combination of the constraints, it (weakly) vanishes, and hence the dynamics is frozen. In 
the quantum theory, the genuine observables commute with the Hamiltonian, and thus they do not evolve. This is the problem of time, which can be addressed using 
relational dynamics by defining observables which provide us relations between different dynamical quantities. An example of the relational 
dynamics is using a matter field $\phi$ as a clock to measure the way gravitational degrees of freedom change. We discuss one such example in the 
spatially flat isotropic and homogeneous spacetime with a massless scalar field in the following.

\subsection{The spatially flat, homogeneous and isotropic spacetime: from Hamiltonian to Friedmann equation}
Let us illustrate the Hamiltonian constraint and how it can be used to obtain classical dynamical equations for the  
the spatially flat isotropic and homogeneous FLRW spacetime. The metric for lapse $N=1$ is given by 
\be\label{metric}
\d s^2 = - \, \d t^2 \, + q_{ab} \d x^a \d x^b = - \d t^2  + a^2 \left(d r^2 + r^2\left(d \theta^2 + \sin^2\theta d \phi^2\right)\right)
\ee
where $a(t)$ is the scale factor of the universe, and $t$ is the proper time. The extrinsic curvature of the spacetime turns out to be $K_{ab} = (\dot a/a) q_{ab}$. 
In the metric variables, the gravitational phase space variables are $a$ and $p_a$, which satisfy: $\{a,p_a\} = 4 
\pi G/3$. If the matter source is a massless scalar field $\phi$, then the matter phase space variables are 
$\phi$ and its conjugate $p_\phi$, which satisfy $\{\phi, p_\phi\} = 1$. Due to the homogeneity of the FLRW spacetime, the spatial 
diffeomorphism constraint is trivially satisfied, and one is 
left with the Hamiltonian constraint, given by
\be\label{cl_cons}
{C_H}^{\rm{(cl)}} = -\f{3}{8 \pi G} \f{p_a^2}{a} + \f{p_\phi^2}{2 a^3}  ~,
\ee
(where we have introduced the superscript ${\mathrm{(cl)}}$ to distinguish the Hamiltonian constraint in the classical theory  from 
the later occurrences). 

Using Hamilton's equations, we can relate the conjugate momenta $p_a$ and $p_\phi$ with $a$ and $\phi$ respectively. It is straightforward 
to see that $p_a$ is related to the time derivative of the scale factor as
\be\label{dota}
\dot a = \{a,{C}_H^{\rm{(cl)}} \} = \f{4 \pi G}{3} \f{\partial}{\partial p_a} \, {C_H}^{\rm{(cl)}} = \f{p_a}{a} ~,
\ee
and similarly $p_\phi = a^3 \dot \phi$. Physical solutions satisfy the vanishing of the Hamiltonian constraint, 
$C_H^{\rm{(cl)}} = 0$, which along with eq.(\ref{dota}) implies
\be
\f{\dot a^2}{a^2} = \f{8 \pi G}{3} \rho, ~~ \mathrm{where} ~~ \rho = \f{p_\phi^2}{2 a^6} ~.
 \ee
Thus, we obtain the Friedmann equation which captures the way the Hubble rate defined as $H = \dot a/a$ varies with the energy density of 
the matter $\rho$. Similarly, it 
is straightforward to obtain the Raychaudhuri equation, and the conservation law using the Hamilton's equation for $p_a$ and $p_\phi$:
\be
\f{\ddot a}{a} = - 4 \pi G (\rho + 3 P), ~~~ \mathrm{and} ~~~ \dot \rho  = - 3 H (\rho + P)
\ee
where $P$ denotes the pressure of the matter component. 

These sets of equations can also be equivalently derived starting from the 
Hamiltonian constraint in the Ashtekar variables. Due to the homogeneity, the matrix valued 
connection $A^i_a$ and the conjugate triad $E^a_i$ can be expressed in terms of the isotropic connections and triads $c$ and $p$, 
which satisfy \citep{abl} 
\be
\{c,p\} = \f{8 \pi G \gamma}{3} ~.
\ee
These variables are related to the metric variables: $|p| = a^2$ and $c = \gamma \dot a$, where the modulus sign arises because the triad 
can have two orientations. Note that the relation between the connection and the time derivative of the scale factor holds only 
in the classical theory, and is not valid when loop quantum gravitational effects are present.

The classical Hamiltonian constraint for the massless scalar field as the matter source turns out to be
\be \label{hc3} {C}_H^{\rm{(cl)}} =  - \f{3}{4\gamma^2} \, b^2
|v|\, + \f{p_\phi^2}{4\pi G |v|} \, = 0 \,, \ee
where $b = c/|p|^{1/2}$ and $v = {\rm sgn}(p) |p|^{\f{3}{2}}/{2\pi G}$, which related to $c$ and $p$ by a canonical transformation. These 
variables are  introduced sine they are very convenient to use in the quantum theory \citep{slqc}.  The sgn(p) is $\pm 1$ depending on the 
orientation of the triad. The variables $b$ and $v$ are conjugate to each other, and satisfy $ \{b,\, v\} = 2\gamma$. Using the Hamiltonian 
constraint, we obtain the Friedmann equation from $\dot v$, and the Raychaudhuri equation using $\dot b$. In the massless scalar field case, 
the equation of state is $w = P/\rho = 1$, and the integration of the dynamical equations 
yields $\rho \propto a^{-6}$  where $a \propto t^{1/3}$. As $t \rightarrow 0$, the scale factor 
approaches zero and the energy density becomes infinite in finite time. The classical trajectories 
can also be obtained relationally in this model. These trajectories are:
\be\label{traj}
\phi = \pm \f{1}{\sqrt{12 \pi G}} \ln \f{v}{v_o} + \phi_o
\ee
where $v_o$ and $\phi_o$ are constants. One trajectory corresponds to the expanding universe which encounters the big bang singularity in 
the past evolution. The other trajectory corresponds to the contracting universe 
which encounters the big crunch singularity in the future evolution. Note that both the branches are 
disjoint. In the classical theory, all the dynamical solutions of this model are singular. Finally we note that these singularities are not a consequence of 
assuming a massless scalar field as matter, but are generic properties of flat FLRW spacetime with matter satisfying weak energy condition, including the inflationary models \citep{bgv}.

\subsection{Curvature invariants, geodesics and the strength of the singularities}
We now discuss some of the key properties of the singularities for the $k=0$ FLRW model. 
In this model, for the massless scalar field, the only singularities are of the big bang/big crunch type, where the spacetime curvature 
invariants diverge. This is 
straightforward to see by computing the Ricci scalar $R$:
\be
R = 6 \left(H^2 + \f{\ddot a}{a} \right) = 8 \pi G (\rho - 3 P) ~.
\ee
For matter with a fixed equation of state $w = P/\rho$, knowing the variation of energy density with time is sufficient to capture the details of the way curvature 
invariants vary in this spacetime. For the case of the massless scalar field, the magnitude of the Ricci scalar diverges in a finite time 
at the big bang/big crunch singularities as $R \propto a^{-6}$. 

At the big bang/big crunch,  geodesic evolution breaks down. The geodesic equations for the flat $(k=0)$ FLRW model is 
\be
(u^{\alpha})^\prime + \Gamma^\alpha_{~ \beta \nu} \, u^\beta u^\nu = 0 ~
\ee
where $u^a = \d x^a/d \tau$. The geodesic equation for the  time coordinate, for massive particles, turns out to be 
\be\label{g1}
t^{\prime \, 2} = \varepsilon + \f{\chi^2}{a^2(t)} 
\ee
where $\chi$ is a constant, and $\varepsilon = 0$ or $1$ depending on whether the particle is massless or massive. Here, $u^x, u^y$ 
and $u^z$ also turn out to be constant. 
The derivative of $t'$ can be obtained using the radial equation $a^2 r^\prime = \chi$, and it turns out be 
\be\label{g2}
\ppt = - a \dot a \, r^{\prime \, 2} = - H \, (t^{\prime \, 2} - \varepsilon) ~.
\ee
The geodesic evolution breaks down when the Hubble rate diverges and/or the scale factor vanishes. At the big bang/big crunch singularity, the Hubble rate diverges 
at the zero scale factor. Hence, geodesics can not be extended beyond these singularities.

Though curvature invariants and geodesics are extremely useful to gain insights on singularities, they do not fully capture all the properties of the singularities. As an example, it is possible that at a spacetime event, the 
curvature invariants may diverge, yet `singularity' may be traversable. For this reason, it is important to have a measure of the strength of the singularities in terms of the 
tidal forces.\footnote{See \citet{clarke-krolak} for a detailed discussion on the strength of singularities.} 
Depending on the strength, the singularities can be classified as strong or weak using conditions formulated by Tipler \citep{tipler} and  
 Kr\'{o}lak \citep{krolak}. In simple terms,  
if the tidal forces are such that an in-falling apparatus with an arbitrary strength is completely destroyed, then the singularity 
is considered strong. Otherwise it is weak. 
For the FLRW spacetime, a singularity is strong a la Kr\'{o}lak if and only if 
\be\label{krolak}
\int^\tau_0 \d \tilde\tau \, R_{ab} u^a u^b ~
\ee
is infinite at the value of $\tau$ when the singularity is approached. Otherwise the singularity is weak. The condition whether the singularity 
is strong or weak is slightly different according to Tipler's criteria \citep{tipler}, which is based on the integral
\be\label{tipler}
\int^\tau_0 \d \tilde\tau \int^{\tilde\tau}_0 \, \d \tilde{\tilde \tau} \, R_{ab} u^a u^b ~.
\ee
If the integral diverges then the singularity is strong, else it is weak. From these two conditions we see that it is possible for a singularity to be strong 
by Kr\'{o}lak's criteria, but weak a la Tipler. Since the weak singularities are not strong enough to cause a complete destruction of  
arbitrary detectors, they are essentially harmless. 
The big bang/big crunch singularities are strong singularities. Recall that at these events geodesic evolution also break down. But, big 
bang/big crunch singularities 
are only one type of possible singularities. Below we discuss various types of singularities, other than the big bang/big crunch which are 
possible in the FLRW spacetime.

\subsection{Different types of singularities in FLRW spacetime}
For matter satisfying a non-dissipative equation of state of a general form $P = P(\rho)$, it is possible that other possible types of 
singularities may arise. Unlike big bang/big crunch singularities, these singularities need not occur at a vanishing scale factor, but 
may even occur at infinite volume. As an example, if we consider a fluid with a fixed equation of state $w < -1$, then the energy density 
 $\rho \propto a^{-3(1 + w)}$ grows as the scale factor increases in the classical theory. In such a case, though the big bang is absent,  
there is a future singularity in finite proper time. Such a singularity can arise in the phantom field models\footnote{See for eg. 
\citet{ssd}}, and is known as the big rip singularity \citep{caldwell}.  Such exotic singularities in general relativity,  
can be classified using the scale factor, energy density $\rho$ and pressure $P$ in the following types:

\noindent
$\bullet$ Type-I singularities: At these events, the scale factor diverges in finite time. The energy density and pressure diverge at these 
events, causing curvature invariants to become infinite. These singularities are also known as big rip singularities.

\noindent 
$\bullet$ Type-II singularities: These events, also known as sudden singularities \citep{sudden}, occur at a finite value of the scale 
factor at a finite time. The energy density is finite, but the pressure diverges which causes the divergence in the curvature invariants. 

\noindent
$\bullet$ Type-III singularities: As in the case of type-II singularities, these singularities also occur at a finite value of the scale factor, but are accompanied by the 
divergence in the energy density as well as the pressure. These singularities are also known as big freeze singularities \citep{mariam}.

\noindent
$\bullet$ Type-IV singularities: Unlike the case of big bang/crunch, and other singularities discussed above, curvature invariants remain 
finite at these singularities. But their 
derivatives blow up \citep{barrowtsagas}. In this sense, such singularities can be regarded as soft singularities.

Above singularities provide a useful set of examples to understand the role of curvature invariants, geodesics and the strength of singularities discussed earlier. 
The type-IV singularities are the weakest, because they  do involve any divergence of the curvature invariants. Though, curvature invariants diverge for 
type-II singularities, they are events beyond which geodesics can be extended \citep{lazkoz}. Further, they turn out to be weak 
singularities. Thus, type-II singularities are important to 
look at more carefully and provide an example that the mere divergence of spacetime curvature does not imply that the event is genuinely singular. Type-I and type-III singularities, 
on the other hand share the properties of big bang/big crunch, in the sense that they are strong\footnote{Big freeze singularities are strong according to Kr\'{o}lak's criteria, but weak by Tipler's criteria. In the following the strength of the big freeze singularity will be referred via the Kr\'{o}lak criteria.} and geodesically inextendible events. 

Finally, let us note that type-I-IV singularities require matter with special energy conditions \citep{visser}. As an example, type-I  
singularities require the violation of the 
null energy condition $(\rho + P) \geq 0$, and type-II singularities require the violation of dominant energy condition $(\rho \pm P) \geq 
0$. However, such singularities are not 
difficult to realize by an appropriate choice of $P(\rho)$ (see for example, \citet{not}).

\section{Loop quantization of cosmological spacetimes: $k=0$ isotropic and homogeneous model}

We now briefly discuss the quantization of the spatially flat, homogeneous and isotropic model with a massless scalar field using the 
techniques of loop quantum gravity. 
This model was the first spacetime to be rigorously quantized in loop quantum cosmology, where the 
physical Hilbert space, inner product and 
physical observables were found and evolution with respect to massless scalar field as a clock was studied in detail using analytical and 
numerical methods \citep{aps1,aps2,aps3}. The model can also be exactly solved \citep{slqc}, thus robustness of various results can be 
verified analytically. 
The loop quantization of the $k=0$ model has been generalized in the presence of spatial curvature \citep{apsv,warsaw,open,szulc}, in the  
presence of cosmological constant \citep{bp,kp,ap} and in the presence of potentials \citep{aps4,cyclic}. The model is very 
useful to understand the loop quantization of anisotropic \citep{chiou,awe2,awe3,we,pswe} and Gowdy models which are the spacetime with 
infinite number of degrees of freedom \citep{gowdy1,gowdy2,gowdy3}. Our discussion follows the analysis of \citet{aps3}, to which an 
interested reader can refer to for further details.

Though one starts with the connection as a gravitational phase space variable at the classical level in the formulation of loop quantum 
gravity, it turns out that at the quantum theory level, the connection has no operator analog. Instead one uses the holonomies of the 
connection as the elementary variable 
for quantization. The holonomies of the isotropic connection $c$ are:
\be
h_i^{(\mu)} \, = \, \cos \left(\f{\mu c}{2}\right) \mathbb{I} + 2  \sin \left(\f{\mu c}{2}\right) \tau_i
\ee
where $\mu$ parameterizes the edge length over which the holonomy is computed, $\mathbb{I}$ is an identity matrix and $\tau_k = - i 
\sigma_i/2$, with $\sigma_i$ as the Pauli spin matrices. To obtain a 
quantum theory at the kinematical level, one finds the representation of the algebra of the elements of the holonomies. These elements are 
the functions $N_\mu := e^{i \mu c/2}$ which satisfy: $\langle N_\mu|N_\mu' \rangle = \delta_{\mu \mu'}$, 
 where $\delta_{\mu \mu'}$ is the Kronecker delta. The kinematical Hilbert space in loop quantum cosmology turns out to be fundamentally different from the  quantization of the same spacetime using metric variables based Wheeler-DeWitt theory \citep{abl}. In contrast to the Wheeler-DeWitt theory, the normalizable states in loop quantum cosmology are a countable sum of $N_\mu$. To understand another difference between loop quantum cosmology and the Wheeler-DeWitt theory, let us consider the states in the triad representation $\Psi(\mu)$ to understand the action of the operators $\hat N_\mu$. In this representation, the triad operators $\hat p$ act multiplicatively:
 \be
\hat p \, \Psi(\mu) = \f{8\pi \gamma \lp^2}{6}\, \mu \Psi(\mu) ~. 
\ee
On the other hand, the action of $N_\mu$ is not differential, as it would be for example in the Wheeler-DeWitt theory, but it is 
translational:
 \be
 \hat N_\chi \, \Psi(\mu) = \Psi(\mu + \chi) ~.
 \ee
This translational action of the holonomies plays an important role in the structure of the  quantum Hamiltonian constraint in loop quantum 
cosmology which is obtained as follows. 
One starts with the classical Hamiltonian constraint in terms of the triads and the field strength of the connection (\ref{hamclassical}), and 
as in the gauge theories expresses the field strength in terms of holonomies over a square loop with area $\bar \mu^2 |p|$. In the gauge theories, the area of the 
loop over which the holonomies are computed can be shrunk to zero, but not in loop quantum cosmology. The reason is tied to the 
underlying quantum geometry in loop quantum gravity which allows the loops to be shrunk to the minimum non-zero eigenvalue of the area operator 
$\Delta = 4 \sqrt{3} \pi \gamma \lp^2$ \citep{awe2}, where $\lp$ denotes the Planck length. This results in $\bar \mu^2 = \Delta/|p|$. 
The functional dependence of $\bar \mu$ on the triads complicates the action of the $\hat N_{\bar \mu}$ operators on the states in the triad 
representation. It turns out that if one instead works in the volume representation, then operators $\widehat{\exp(i \lambda b)}$ (where $\lambda = \sqrt{\Delta}$) shift states in uniform steps in volume. 

The resulting Hamiltonian constraint in loop quantum cosmology in the volume representation can then be written as,
\be \label{cons1}
\partial_\phi^2 \, \Psi(\nu,\phi) \, = \, 3 \pi G \, \nu \, \f{\sin \lambda b}{\lambda} \nu \, \f{\sin \lambda b}{\lambda} \, \Psi(\nu,\phi) \, =: \, - \, \Theta \Psi(\nu,\phi) \\
\ee
where  $\nu = v/\gamma\hbar$, and $\Theta$ is a difference operator:
\be\label{thetacons}
\Theta \Psi(\nu, \phi) \, := \, - \f{3 \pi G}{4 \lambda^2} \, \nu \left((\nu + 2 \lambda) \Psi(\nu + 4 \lambda) - 2 \nu \Psi(\nu,\phi) + (\nu - 2 \lambda) \Psi(\nu - 4 \lambda)\right) ~.
\ee
In contrast, in the Wheeler-DeWitt theory the quantum Hamiltonian constraint turns out to be:
\be\label{wdwcons}
 \p_\phi^2 \ul{\Psi}(z,\phi) = 12\pi G\, \partial_z^2 \ul{\Psi}(z,\phi) =: - \ul\Theta  \, \ul\Psi(z,\phi)\,  
 \ee
 where $z = \ln a^3$. Note that in loop quantum cosmology, the difference operator arises due to the underlying quantum geometry. However, 
 in the Wheeler-DeWitt theory the underlying geometry is a continuum, and the quantum Hamiltonian constraint is a differential operator. In 
the 
 limit of large volume, the discrete quantum operator $\Theta$ is very well approximated by the continuum differential operator $\ul \Theta$. 
 Since in this model, the large volumes correspond to the regime where spacetime curvature is small, one concludes that the discrete 
 quantum geometry can be approximated by the continuum differential geometry when the gravitational field becomes weak. 
 
In the quantum theory, the scalar field $\phi$ plays the role of internal time with respect to which evolution of observables can be studied. 
These observables are the volume of the universe at time $\phi$, and the momentum of the scalar field (which is a constant of motion). These 
observables are self-adjoint with respect to the inner product:
\be
\langle \Psi_1| \Psi_2\rangle \, = \, \sum_\nu \, \bar \Psi_1(\nu,\phi_o) |\nu|^{-1} \Psi(\nu_2,\phi_o) ~.
\ee 
With the availability of the inner product and the self-adjoint observables, one can extract physical predictions in the quantum theory. 
One can choose a state, such as a sharply peaked state, at late times on a classical trajectory and evolve the state numerically using (\ref{cons1}). The expectation values of the observables can then be computed and compared with the classical trajectory. In Fig. 1, we 
show the results from such an evolution for loop quantum cosmology and the Wheeler-DeWitt theory by considering a semi-classical 
state in both the theories at large volumes. If the state is chosen peaked in an expanding branch, we evolve it backwards towards the big 
bang using $\phi$ as a clock. In the 
Wheeler-DeWitt theory, the resulting expectation values of the volume observable (depicted by ${\rm{WDW}}_{\rm{exp}}$ in Fig. 1) matches with the classical trajectory throughout the 
evolution. Similarly, the state can be chosen in the contracting branch and evolved in future towards the big crunch. In this case too, the 
Wheeler-DeWitt theory (with expectation values depicted by ${\rm{WDW}}_{\rm{cont}}$ in Fig. 1) turns out to be  in agreement with general 
relativity at all the scales. The Wheeler-DeWitt theory, though a quantum theory of 
spacetime, yields the classical description in the entire evolution. The expanding and contracting branches in the Wheeler-DeWitt theory 
encounter big bang and big crunch singularities respectively, as in general relativity. One can also compute the quantum probability for 
the occurrence of singularity in the Wheeler-DeWitt theory, which turns out to be unity \citep{craig-singh1}.

\begin{figure}
  \centerline{\includegraphics[angle=270,width=12cm]{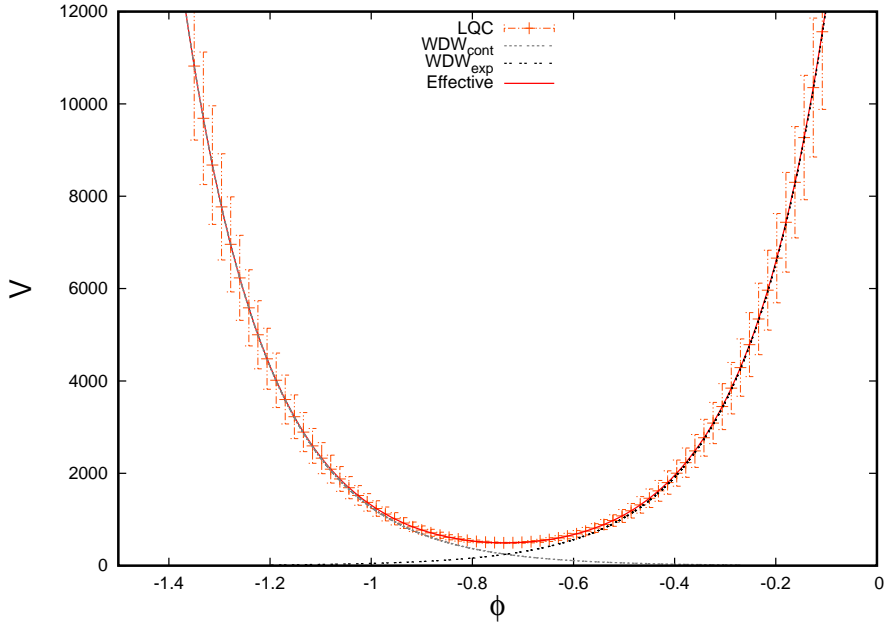}}
  \caption{This plot shows the expectation values of the volume observable plotted against `time' $\phi$ in loop quantum cosmology and the 
Wheeler-DeWitt theory. In loop quantum cosmology, the quantum evolution results in a bounce. In contrast, in the Wheeler-DeWitt theory the 
expanding and the contracting branches are disjoint and singular. The loop quantum evolution and the Wheeler-DeWitt evolution are in 
 agreement when the spacetime curvature is very small. We also show the trajectory obtained from the effective Hamiltonian 
constraint in loop quantum cosmology. As can be seen, the latter provides an excellent approximation to the 
underlying quantum dynamics.}
  \end{figure}

In loop quantum cosmology, we find that the state does not encounter the big bang. Rather, it bounces at a certain volume determined by 
the field momentum $p_\phi$, a constant of motion, on which the initial state is peaked. It turns out that if the state is sharply peaked 
then irrespective of the choice of $p_\phi$, the volume (or the scale factor) of the universe bounces when the energy density of the field 
reaches a value $\rho_{\rm{max}} \approx 0.41 \rho_{\rm{Pl}}$ \citep{aps3}. As we can see from Fig. 1, at large values of volume the 
evolution in loop quantum cosmology and the Wheeler-DeWitt theory are in excellent agreement. It is only in the small neighborhood of the 
bounce that there are significant departures 
between the two theories occur. The quantum gravitational effects in loop quantum cosmology bridge the two singular trajectories in the 
classical theory. In Fig. 1, we also show a curve which corresponds to a trajectory derived from an effective Hamiltonian constraint in 
loop quantum cosmology (discussed in the next section), which turns out to be in excellent agreement with the quantum theory at all the scales.

At this point it is natural to ask whether the results of quantum bounce are robust. This question has been answered in several ways. It 
turns out that the model we studied here can be solved exactly in the $b$ representation \citep{slqc}. One can then compute the expectation 
values of volume observable and energy density without resorting to the numerical simulations. It turns out that for all the states in the 
physical Hilbert space, 
the expectation values of volume have a non-zero minima, and those of the energy density have a supremum equal to $0.41 \rho_{\rm{Pl}}$. 
Thus, the results of the exactly soluble model are in complete agreement with the numerical simulations. Further, the quantum probability 
for the bounce to occur for generic states turns out to be 
unity \citep{craig-singh2}. 
Another robustness check comes from rigorous numerical tests \citep{ps12,khanna}, including numerical simulations with states which may not be sharply peaked. Recent numerical studies 
confirm that the quantum bounce occurs for various types of states including those which may have very large quantum fluctuations and 
non-Gaussian properties \citep{dgs2,squeezed}. In fact, the profile of the state 
is almost preserved and relative fluctuations are tightly constrained across the bounce, in agreement with the analytical results from the 
exactly soluble model \citep{cs1,triangle,cm1}. 

One of the important issues in a quantization is that of ambiguities. Are different consistent loop quantizations of FLRW model possible? 
Is it possible that the quantum Hamiltonian constraint be a uniform discrete equation in a geometric variable such as area rather than volume? It is interesting to note that by carefully considering mathematical consistency of such alternatives, and demanding that the resulting theory should lead to physics independent of the underlying fiducial structure, one is uniquely led to the loop quantization which is discussed above \citep{cs2}. Similar conclusions have been reached for the Bianchi and Kantowski-Sachs spacetimes \citep{cs3,pswe,js1}.

\section{Effective spacetime description of the isotropic loop quantum cosmology and the generic resolution of singularities}
In the previous section, we found that the quantization of the Hamiltonian constraint in loop quantum cosmology results in a quantum  
difference equation with uniform steps in 
volume. We also discussed that when gravitational field becomes weak, the quantum difference equation leads to the continuum general relativistic description. 
An interesting question is whether there exists an effective continuum spacetime description in loop quantum cosmology which reliably 
captures the underlying physics. If such a description is available, then it can be an important tool to understand the physical 
implications of loop quantum cosmology.  Using the geometrical formulation of the 
quantum theory \citep{schilling}, an effective Hamiltonian constraint in loop quantum cosmology can be obtained for states which are peaked 
on 
a classical trajectory at late times, i.e. for universes which grow to a macroscopic sizes \citep{jw,vt}. The 
dynamics obtained from the 
effective Hamiltonian  turns out to be in excellent agreement with the underlying quantum dynamics for the sharply peaked states 
\citep{aps3,dgs2}, which can be seen from the comparison of the quantum evolution and effective trajectory in Fig. 1.

The effective Hamiltonian constraint for the $k=0$ isotropic and homogeneous spacetime is:
 \be \label{effham}
{C}^{\mathrm{(eff)}}_H = - \f{3}{8 \pi G \gamma^2} \, \f{\sin^2(\lambda b)}{\lambda^2} V +
\, \rho V ~ ~,
\ee
where $b$ and $V = |p|^{3/2}$ are the conjugate variables satisfying $\{b,V\} = 4 \pi G \gamma$, and $\lambda = \sqrt{\Delta}$. The 
vanishing of the 
Hamiltonian constraint leads to 
\be\label{eq:sin_rho}
\f{\sin^2(\lambda b)}{\gamma^2 \lambda^2} = \f{8 \pi G}{3} \, \rho ~.
\ee
Note that the right hand side of the above equation has the same form as that of the classical Friedmann equation, but the left hand side is not the Hubble rate. 
We can rewrite the above equation in terms of the Hubble rate $H = \dot a/a = \dot V/3V$, by using the Hamilton's equation: 
\be\label{vdot}
\dot V =  - 4 \pi G \gamma \f{\partial}{\partial b} {C}^{\mathrm{(eff)}}_H = \f{3}{\gamma} \, \f{\sin(\lambda b)}{\lambda} \, \cos(\lambda b) \, V ~.
\ee
Substituting the above equation in eq.(\ref{eq:sin_rho}) and using a trigonometric identity one obtains the modified Friedmann equation in 
loop quantum cosmology \citep{aps3,ps06}\footnote{A similar modification to the Friedmann equation arises for extra dimensional brane world 
models with two time-like dimensions \citep{ss}.}:
\be \label{fried}
H^2 = \f{8 \pi G}{3} \, \rho \left(1 - \f{\rho}{\rcr}\right)
\ee
where $\rcr$ is a constant determined by the quantum discreteness $\lambda$,
\be\label{rhocrit}
\rcr  = 3/(8 \pi G \gamma^2 \lambda^2) \approx 0.41 \rho_{\mathrm{Pl}} ~.
\ee
One can similarly derive the modified Raychaudhuri equation in loop quantum cosmology by taking the time derivative of eq.(\ref{vdot}) and using the Hamilton's equation 
$\dot b = \{b, {C}^{\mathrm{(eff)}}_H\}$, which yields
\be\label{rai}
\f{\ddot a}{a} = - \f{4 \pi G}{3} \, \rho \, \left(1 - 4 \f{\rho}{\rcr} \right) - 4 \pi G \, P \, \left(1 - 2  \f{\rho}{\rcr} \right) . ~
\ee
Combining the modified Friedmann and Raychaudhuri equations, it is straightforward to see that one recovers the conservation law: $\dot \rho 
+ 3 H \, (\rho + P) = 0$. 

Let us now analyze some of the main properties of the modified Friedmann and Raychaudhuri equations in loop quantum cosmology. We first 
note from eq.(\ref{fried}) that the physical solutions have an upper bound on the energy density given by $\rcr$, where the Hubble rate 
vanishes. This upper bound coincides with the value of bounce density in the numerical simulations with the states which are sharply 
peaked at late times on the classical trajectory. If the states have very large quantum fluctuations, then there exist departures between 
the quantum evolution and effective dynamics \citep{dgs2}. In that case, the bounce density in the quantum theory is always less 
than $\rcr$. The modified Friedmann equations in loop quantum cosmology result lead to the classical Friedmann and Raychaudhuri equations 
when the quantum discreteness $\lambda$ vanishes. In this case, the maximum of the energy density $\rcr$ becomes infinite and the classical 
singularity is recovered. As mentioned earlier, the non-singular modified Friedmann equations in loop quantum cosmology results in a very rich phenomenology for 
the background dynamics as well as the cosmological perturbations. For a review of these applications, we refer the reader to \citet{asrev}.

Unlike in the classical theory where Hubble rate diverges, such as at the big bang singularity, in loop quantum cosmology the Hubble rate is 
bounded in the entire evolution. Its maximum value is  
\be\label{hmax_iso}
|H|_{\mathrm{max}} = \left(\f{1}{\sqrt{3} \, 16 \pi G \hbar \gamma^3}\right)^{1/2} ~
\ee
which is reached at $\rho = \rcr/2$. A consequence of the bounded Hubble rate is that except for the events where the scale factor either 
vanishes or diverges, the geodesic equations (\ref{g1}) and (\ref{g2}) never break down. But what if the singularity occurs 
with a vanishing or a diverging scale factor? It turns out that evolution results in either of these possibilities, then the universe in 
loop quantum cosmology behaves asymptotically as the classical deSitter universe with a rescaled value of the cosmological constant. In the 
classical theory, geodesics can be extended in the case of the deSitter universe, and thus these  points are not problematic for geodesic 
evolution. Finally, one can rigorously show that the spatially flat isotropic and homogeneous spacetime in effective dynamics of loop 
quantum cosmology is geodesically complete \citep{ps09}.

It is interesting to note that curvature invariants remain bounded irrespective of the choice of matter. As an example, the Ricci scalar turns out to be 
The Ricci curvature invariant turns out to be 
\be
\label{Ricci}
R =  8 \pi G \rho \, \left(1 - 3 w + 2 \f{\rho}{\rcr} \left(1 + 3 w \right) \right) .
\ee
where $w = P/\rho$. The Ricci scalar, and similarly other curvature invariants, are bounded for all the events where the energy density 
diverges. Thus, big bang/big crunch, bug rip and big freeze singularities are avoided in loop quantum cosmology. Thus all the 
strong singularities of the classical FLRW model are absent in loop quantum cosmology \citep{ps09}.

 However, the Ricci scalar can diverge when $|w| \rightarrow \infty$, which happens for the type-II singularities. Such `soft singularities' 
have  been shown to exist 
in loop quantum cosmology \citep{portsmouth,ps09,psfv}, but these are harmless because they turn out to be weak singularities in the 
effective spacetime (as in the classical general relativity). This is straightforward to see from the   Kr\'{o}lak (\ref{krolak}) and Tipler 
(\ref{tipler}) conditions for the existence of strong singularities. As an example, the integrand in Kr\'{o}lak and Tipler's conditions for 
the null geodesics can be written as 
\be
R_{ij} u^i u^j = 8 \pi G (\rho + P) \f{\chi^2}{a^2} \left(1 - 2 \f{\rho}{\rcr}\right)
\ee
which yields the finite result for the integrals in (\ref{krolak}) and (\ref{tipler}) for type-II singularities. Thus, one finds that the 
strong singularities are completely avoided in the effective spacetime description of loop quantum cosmology for the spatially flat, 
isotropic and homogeneous model and only weak singularities can exist. The geodesic evolution never break down for an arbitrary choice of  
matter. These results are in a striking contrast to the classical FLRW model where unless the universe is deSitter, in general the spacetime 
is geodesically incomplete and strong singularities exist either in the past or the future evolution. 

The robustness of the results of the spatially flat isotropic model has been studied for the $k = \pm 1$ models using a phenomenological 
equation of state allowing all the types of singularities \citep{psfv}. This analysis confirms that along with the big bang/big crunch, all 
type-I and type-III singularities are resolved generically.  As in the $k=0$ model,  
quantum gravitational effects ignore the harmless weak singularities. These results have also been generalized to include the anisotropies 
which we discuss in the following section.


\section{Generic resolution of singularities in presence of anisotropies}
So far we have focused our discussion only on the isotropic models where we demonstrated the resolution of singularities for arbitrary matter 
in the effective spacetime description of loop quantum cosmology. We now briefly discuss the way these results generalize to the 
inclusion of anisotropies in the spacetime. For this we consider the Bianchi-I spacetime which has a spacetime metric:
 \be\label{metric_b1}
d s^2 = -\, d t^2 + a_1^2 \, d x^2 + a_2^2 d y^2 + a_3^2 d z^2 ~.
\ee
 As in the homogeneous isotropic model, in the Bianchi-I spacetime the Ashtekar variables can be expressed 
in terms of symmetry reduced connections  and triads: $c_i$ and triads $p_i$, which satisfy $\{c_i, p_j\} = 8 \pi G \gamma \delta_{ij}$. The triads 
are kinematically related to the scale factors as,
 \be\label{triadsf}
|p_1| = a_2 a_3, ~~~ |p_2| =  a_1 a_3, ~~~ |p_3| = a_3 a_1 ~.
\ee
Due to homogeneity, the diffeomorphism constraint is satisfied and one is left only with the Hamiltonian constraint, 
which in terms of $c_i$ and $p_i$ is:
\be\label{clH}
{C}^{\mathrm{(cl)}}_H = -\f{1}{8 \pi G \gamma^2 V}{(c_1 p_1 \, c_2 p_2 + c_3 p_3 \, c_1 p_1 + c_2 p_2 \, c_3 p_3)} + \rho V ~ ,
\ee
where $\rho$ denotes the energy density of the matter which is assumed to be of the vanishing anisotropic stress, i.e. it satisfies 
$\rho(p_1,p_2,p_3) = \rho(p_1 p_2 p_3)$. As in the case of the isotropic model, we can use the Hamilton's equations, along with the 
Hamiltonian constraint, to obtain the dynamical equations. One of these equations is the generalized Friedmann equation, 
\be\label{fried_cl}
H^2 = \f{8 \pi G}{3} \rho + \sigma^2 ~
\ee
where $H$ denotes the mean Hubble rate: $H = (H_1 + H_2 + H_3)/3 = \dot V/3V$, and $\sigma^2$ is the shear scalar:
\be\label{sigmadef}
\sigma^2 = \f{1}{3} \left((H_1 - H_2)^2 + (H_2 - H_3)^2 + (H_3 - H_1)^2\right) ~.
\ee
In the classical theory, $\sigma^2 \propto a^{-6}$. An important implication of this behavior is that unless the equation of state of 
matter $w$ is greater than or equal to unity, the anisotropies dominate near the singularities. As discussed earlier, the big bang/big crunch 
singularities  can have various structures. These are: 
a  barrel, where one of the scale factor takes a finite values and the other two scale factors vanish; a cigar, where one of the 
scale factors diverges, and the other two vanish; a pancake, where one of the scale factors vanish and the other two take a finite value; and 
a point, where all the scale factors vanish. The point like singularity is the isotropic singularity. At these singularities, the directional Hubble rates $H_i$, $\rho$ and $\sigma^2$ diverge and the 
geodesic evolution breaks down. This can be seen from analyzing the following geodesic equations in the Bianchi-I spacetime:
\be \label{geod1}
x^{\prime \prime} = -2 x' t' H_1, ~~ y^{\prime \prime} = -2 y' t' H_2, ~~ z^{\prime \prime} = - 2 z' t' H_3, ~~ t^{\prime \prime} = - a_1^2 H_1 \,x'^2 - a_2^2 H_2 \,y'^2 - a_3^2 H_3 \,z'^2
\ee
and
\be \label{geod2}
x' = \f{k_x}{a_1^2}, ~~~  y' = \f{k_y}{a_2^2}, ~~~ z' = \f{k_z}{a_3^2}, ~~~ t' = \left(\f{k_x^2}{a_1^2} + \f{k_y^2}{a_2^2} + \f{k_z^2}{a_3^2}\right)^{1/2} ~,
\ee
where $k_i$ are constants.  From the above equations we find that geodesics break down when any of the scale factors vanishes or the 
directional Hubble rate diverges. At these events, the curvature invariants blow up. An example of the curvature invariant is the Ricci scalar, whose expression turns out to be 
\be\label{ricci}
R = 2 \left(H_1 H_2 + H_2 H_3 + H_3 H_1 + \sum_{i=1}^3 \f{\ddot a_i}{a_i} \right) ~.
\ee
We can see that at the big bang/big crunch singularities where the directional Hubble rates and $\ddot a_i/a_i$ diverge, the Ricci scalar 
becomes infinite. The same is the fate of the other curvature invariants, such as the Kretschmann and the square of the Weyl curvature. 
These curvature invariant diverging events are  strong curvature singularities. Note that for the anisotropic 
spacetimes, the Tipler and Kr\'{o}lak's conditions also involve integrals over the Weyl curvature components, which should be finite for 
the singularity to be weak.

Let us now discuss the fate of the singularities in the effective spacetime description of loop quantum cosmology. The loop quantization of 
the Bianchi-I model has been rigorously performed by \citet{awe2}. Following the strategy for the quantization of the isotropic spacetimes 
in loop quantum cosmology, the resulting quantum Hamiltonian constraint is a difference operator which turns out to be non-singular. As in 
the case of the isotropic model, the quantum Hamiltonian constraint in the   Bianchi-I model leads to an effective Hamiltonian 
given by \citep{cv},
\be\label{effham_bianchi}
C_H^{(\rm{eff})} =  - ~\f{1}{8 \pi G \gamma^2 V}\left(\f{\sin(\bar \mu_1 c_1)}{\bar \mu_1} \f{\sin(\bar \mu_2 c_2)}{\bar \mu_2}  p_1 p_2 + 
\mathrm{cyclic} ~~ \mathrm{terms}\right) ~~ + ~~ \rho V \\
\ee
where $\bar \mu_i$ are: 
\be \label{mub1}
\bar \mu_1 = \lambda \sqrt{\f{ p_1 }{p_2 p_3}}, ~~~ \bar \mu_2 = \lambda \sqrt{\f{p_2}{p_1 p_3}}, ~~~ \mathrm{and} ~\bar \mu_3 = \lambda \sqrt{\f{p_3}{p_1 p_2}} ~ ,
\ee
and the orientation of the triads is chosen to be positive without any loss of generality. The quantum discreteness is captured by $\lambda$, whose square is the minimum 
area, $\Delta = 4 \sqrt{3} \pi \gamma \lp^2$, to which loops are shrunk while computing holonomies in loop quantum cosmology. An immediate 
consequence of the quantum discreteness is the boundedness of the energy density which follows from the vanishing of the Hamiltonian 
constraint 
\be
\label{rhob1} \rho=\frac{1}{8 \pi G \gamma^2 \lambda^2}\left(\sina \sinb + \mbox{cyclic terms}\right) \leq \f{3}{8 \pi G \gamma^2 \lambda^2}  \approx0.41 \rho_{\rm Pl}
\ee
 Therefore, in contrast to the classical Bianchi-I model, the energy density can never diverge in the loop quantized Bianchi-I spacetime, and 
 interestingly, the value of the maximum energy density turns out to be the same as in the isotropic model.

Using Hamilton's equations we can compute the time derivatives of the triads and obtain the expressions for the directional Hubble rates. These 
turn out to be universally bounded, and so is the mean Hubble rate which has a maximum equal to the maximum Hubble rate in the isotropic model (\ref{hmax_iso}):
\be\label{hib1_max}
H_{\mathrm{(max)}} = \f{1}{2 \gamma \lambda} ~.
\ee
From the directional Hubble rates, it is straightforward to find the anisotropic shear scalar:
\ba
\label{shearb1} \sigma^2 &=& \nonumber \frac{1}{3 \gamma^2 \lambda^2 }\Bigg[ (\cosb(\sina+\sinc)- \cosa(\sinb+\sinc))^2  \\ && +~\mbox{~cyclic terms}~\Bigg] ~,
\ea
which is bounded above by 
\be
 {\sigma^2}_{\rm max}=\frac{10.125}{3 \gamma^2 \lambda^2} .
\ee
Thus, the underlying quantum geometric effects incorporated via the minimum area eigenvalue $\lambda^2$, bind the mean Hubble rate and the shear scalar  in Bianchi-I loop quantum cosmology \citep{cs3,ps11}. If $\lambda$ vanishes, the discrete quantum geometry is replaced by the  classical differential geometry, and the magnitude of the above physical quantities have no upper bound. As we discuss below, 
the boundedness of the mean Hubble rate and the shear scalar has important consequences for the fate of the geodesics and the possibility of 
the existence of strong singularities in the effective spacetime of Bianchi-I model in loop quantum cosmology. 

Note that as we discussed in the case of the isotropic model, even though the energy density, mean Hubble rate and the shear scalar are bounded, the curvature invariants can still diverge for certain choices of equation of states. Analysis of the curvature invariants shows that they are generally bounded but potential divergences can arise if there exists a physical solution for which at a finite value of $\rho, \theta$ and $\sigma^2$,  pressure diverges and/or the mean volume  vanishes \citep{ps11}. None of the known singularities in general relativity satisfy these conditions. Hence, in loop quantum cosmology all of the classical events where curvature invariants diverge are avoided.  Does the existence of these potential curvature invariant divergences signal a strong singularity? The answer is tied to whether the divergence occurs due to pressure becoming infinite or the vanishing of one or more scale factors. If the curvature invariants become infinite  because of the divergence in the pressure then 
the event is a weak singularity. 
However, if the curvature invariants diverge because one of the scale factors vanishes, then the above potential event is a strong singularity. 
It is important to emphasize that the latter events must occur at a finite value of energy density, mean Hubble rate and the shear scalar. 
In general relativity, there are no known singularities which satisfy these conditions in the Bianchi-I model, and existence of these events 
is only a potential mathematical possibility.

Finally, let us discuss the geodesic evolution in the effective spacetime of Bianchi-I model in loop quantum cosmology. Geodesic equations (\ref{geod1}) and (\ref{geod2}) break down when any of the directional Hubble rates diverge and/or the scale factors $a_i$ vanishes. 
In the classical theory, at the big bang/big crunch singularity at least one of the scale factors always vanish, and one the directional Hubble rate diverges. For the other strong singularities which are of big rip and big freeze type, the scale factors remain finite, however the directional Hubble rates diverge. Thus geodesic evolution in classical Bianchi-I model breaks down at these singularities. In contrast, in loop quantum cosmology all the 
classical singularities accompanied by a divergence in the directional Hubble rates and the vanishing of the directional scale factors are forbidden, because the former are universally bound by $H_{i ~ \rm{max}} = 3/(2 \gamma \lambda)$. The fate of the geodesics for the 
mathematically possible  cases where curvature invariants may diverge in loop quantum cosmology depends on the behavior of the 
directional scale factors. If the curvature invariants diverge when one of the scale factors vanish at a finite energy density, mean Hubble rate and the shear scalar, then the geodesic equations will break down. However, if the curvature invariants diverge because the pressure becomes infinite, such as in the analog of sudden singularities in the Bianchi-I model, then the geodesics can be extended beyond such events \citep{ps11}.  We expect similar results to hold in more general spacetimes, such as the Gowdy models where the above results on boundedness of curvature invariants have already been extended \citep{madridgowdy}.

\section{Summary}
  
The existence of  strong singularities in  classical spacetimes,  signals that a more complete description of the universe will include features of quantum properties of spacetime. These properties which will capture the quantum geometric nature of spacetime are expected to provide insights 
to many fundamental questions concerning the birth of our universe, the emergence of classical spacetime and the resolution of singularities. In this manuscript, we 
gave an overview of some of the developments in the framework of loop quantum cosmology, where one uses the techniques of loop quantum gravity to quantize cosmological spacetimes. 
Attempts to quantize cosmological models date back to the Wheeler-DeWitt theory, however, unlike Wheeler-DeWitt quantum cosmology where resolution of singularities was 
generally absent, and if present, required very special boundary conditions, in loop quantum cosmology, resolution of big bang singularity has been found to be a robust feature of all the loop quantized  spacetimes. As an example, if one considers a sharply peaked state at late times on a classical trajectory in a loop quantized spatially flat homogeneous and isotropic spacetime sourced with a massless scalar field, then such a state follows the classical trajectory in the backward evolution for a very long time, all the way close to the Planck scale, but then bounces to a pre-bounce branch without encountering the big bang singularity \citep{aps1,aps2,aps3}.  The existence of bounce has been demonstrated to be a robust feature of various spacetimes in loop quantum cosmology. Investigations of the anisotropic models reveal that the evolution is non-singular, and 
indicate the existence of bounce at the quantum level. These results also extend to the Gowdy models which have infinite degrees of freedom. The underlying loop quantum dynamics 
can be accurately captured by a continuum non-classical effective description which mimics the quantum evolution very accurately even at the bounce.

The existence of bounce in loop quantum cosmology is a direct ramification of the underlying quantum geometry. The discreteness in geometry bounds the energy density, Hubble rate and anisotropic shear, leading to finite curvature invariants throughout the evolution for all the types of matter with an equation of state which lead to a  divergence in the energy density in the classical theory. All of the matter models fall in to this category, except the one with an exotic equation of state,  allowing a divergence in pressure at a finite energy density. However, the divergence in curvature invariants does not signal a genuine singularity. Using effective spacetime description, one finds that the events where the curvature invariants diverge in the classical theory, are either completely eliminated in loop quantum cosmology or tun out to be weak singularities \citep{ps09,ps11}. These singularities are harmless because detectors with sufficient strength can propagate across them. Analysis of the geodesic 
equations signals that the effective spacetime in the spatially flat isotropic model is geodesically complete. For the Bianchi-I model, geodesic evolution does not break down for geodesically inextendible events in the classical theory. Investigations on the behavior of geometric scalars in  other Bianchi models \citep{gs1,pswe}, linearly polarized hybrid Gowdy spacetimes  \citep{madridgowdy} and Kantowski-Sachs spacetime \citep{js1} indicates that these results may hold in a more general setting.

Thus, in contrast to the 
classical theory where singularities are a generic feature, there is a growing evidence in loop quantum cosmology that singularities may be absent. 
An important question in quantum cosmology is whether the results obtained in the homogeneous spacetimes can be trusted in a 
more general setting. As far as the issue of singularity resolution is concerned, there is a strong evidence from the numerical studies \citep{berger,garfinkle} of the Belinskii-Khalatnikov-Lifshitz conjecture \citep{bkl}, that  near the singularities the structure of the spacetime is not determined by the spatial derivatives, but by the time derivatives, and the approach to the singularity can be described via homogeneous cosmological models. Thus one can expect that singularity resolution in homogeneous models would capture some aspects of the singularity resolution in more general spacetimes. Recent work in  relating the loop quantization of spatially flat isotropic and homogeneous spacetime to that of the Bianchi-I model, also provides useful insights on the role of symmetry reduction \citep{awe2}. These results provide evidence that 
quantization of homogeneous models may reliably capture the nature of quantum spacetime in general near the classical singularities, and to some 
extent alleviate the concerns about the role of symmetries.  One can hope that future work in this direction will keep providing important insights on 
the fundamental issues both in 
general relativity and quantum gravity. In particular, one hopes that future investigations on the lines discussed in this manuscript may reveal some important clues to a non-singularity theorem in quantum gravity.

\section*{Acknowledgements}
The author thanks the Astronomical Society of India for awarding the 2010 Vainu Bappu gold medal. The author is grateful to Abhay Ashtekar, David Craig, Alejandro Corichi, Peter Diener, Brajesh Gupt, Anton Joe, Miguel Megevand,  Jorge Pullin, Edward Wilson-Ewing and Kevin 
Vandersloot for many insightful discussions. This work is supported by NSF grants PHY1068743, PHY1404240 and by a grant from the  John Templeton 
Foundation. The opinions expressed
in this publication are those of authors and do not necessarily reflect the views of the John Templeton Foundation.





\begin{thebibliography}{}

\bibitem[Agullo, Ashtekar \& Nelson (2013a)]{aan1} Agullo I., Ashtekar A., Nelson W., 2013a, PhRvD, 87, 043507 

\bibitem[Agullo, Ashtekar \& Nelson (2013b)]{aan2} Agullo I., Ashtekar A., Nelson W., 2013b, CQGra., 30, 085014 

\bibitem[Ashtekar(1986)Ashtekar]{abhay} Ashtekar A., 1986, PhRvL, 57, 2244 
  
\bibitem[Ashtekar \& Bojowald(2006)]{ab_bh} Ashtekar A., Bojowald M., 2006, CQGra, 23, 391  

\bibitem[Ashtekar, Bojowald \& Lewandowski(2003)Ashtekar, Bojowald \& Lewandowski]{abl} 
         Ashtekar A., Bojowald M., Lewandowski J., 2003, Adv. Theor. Math. Phys., 7, 233

\bibitem[Ashtekar, Corichi, \& Singh(2008)Ashtekar, Corichi, \& Singh]{slqc} 
         Ashtekar A., Corichi A., Singh P., 2008, PhRvD, 77, 024046 

\bibitem[Ashtekar \& Lewandowski(2004)Ashtekar \& Lewandowski]{report} 
         Ashtekar A., Lewandowski J., 2004, CQGra, 21, 53 

\bibitem[Ashtekar et al.(2006a)Ashtekar, Pawlowski, \& Singh]{aps1} 
         Ashtekar A., Pawlowski T., Singh P., 2006a, PhRvL, 96, 141301 

\bibitem[Ashtekar et al.(2006b)Ashtekar, Pawlowski, \& Singh]{aps2} 
         Ashtekar A., Pawlowski T., Singh P., 2006b, PhRvD, 74, 084003 

\bibitem[Ashtekar et al.(2006c)Ashtekar, Pawlowski, \& Singh]{aps3} 
         Ashtekar A., Pawlowski T., Singh P., 2006c, PhRvD, 73, 124038 

\bibitem[Ashtekar et al.(2014)Ashtekar, Pawlowski, \& Singh]{aps4} 
         Ashtekar A., Pawlowski T., Singh P., 2014, To appear 

\bibitem[Ashtekar et al.(2007)Ashtekar et al.]{apsv} 
         Ashtekar A., Pawlowski T., Singh P., Vandersloot K., 2007, PhRvD, 75, 024035 
  
\bibitem[Ashtekar \& Schilling(1999)Ashtekar \& Schilling]{schilling} 
         Ashtekar A., Schilling T.~A., 1999, On Einstein's Path, essays in honor of Engelbert Schucking, 
         Ed. Alex Harvey, Springer-Verlag, p 23

\bibitem[Ashtekar \& Singh(2011)Ashtekar \& Singh]{asrev} 
         Ashtekar A., Singh P., 2011, CQGra, 28, 213001 

\bibitem[Ashtekar \& Sloan(2011)Ashtekar \& Sloan]{as1} 
         Ashtekar A., Sloan D., 2011, GReGr, 43, 3619 

\bibitem[Ashtekar \& Sloan(2010)Ashtekar \& Sloan]{as2} 
         Ashtekar A., Sloan D., 2010, PhLB, 694, 108 
  
\bibitem[Ashtekar \& Wilson-Ewing(2009a)Ashtekar \& Wilson-Ewing]{awe2} 
         Ashtekar A., Wilson-Ewing E., 2009a, PhRvD, 79, 083535 
 
\bibitem[Ashtekar \& Wilson-Ewing(2009b)Ashtekar \& Wilson-Ewing]{awe3} 
         Ashtekar A., Wilson-Ewing E., 2009b, PhRvD, 80, 123532  

\bibitem[Barrau et al.(2014)]{barrau} 
         Barrau A., Cailleteau T., Grain J., Mielczarek J., 2014, CQGra., 31, 053001 

\bibitem[Barrow(2004)Barrow]{sudden} Barrow J.~D., 2004, CQGra, 21, 5619 

\bibitem[Barrow \& Tsagas (2005)]{barrowtsagas} Barrow J.D., Tsagas C.G., 2005, CQGra., 22, 1563
  
\bibitem[Belinskii et al.(1970)Belinskii, Khalatnikov, \& Lifshitz]{bkl} 
         Belinskij V.~A., Khalatnikov I.~M., Lifshits E.~M., 1970, AdPhy, 19, 525 
  
\bibitem[Bentivegna \& Pawlowski(2008)Bentivegna \& Pawlowski]{bp} 
         Bentivegna E., Pawlowski T., 2008, PhRvD, 77, 124025  
 
\bibitem[Berger(2002)Berger]{berger} 
         Berger B.~K., 2002, LRR, 5, 1  
 
\bibitem[Brizuela et al.(2012)Brizuela, Cartin, \& Khanna]{khanna} 
         Brizuela D., Cartin D., Khanna G., 2012, SIGMA, 8, 1  

\bibitem[Brizuela et al.(2010)Brizuela, Mena Marug{\'a}n, \& Pawlowski]{gowdy3} 
         Brizuela D., Mena Marug{\'a}n G.~A., Pawlowski T., 2010, CQGra, 27, 052001 

\bibitem[Bojowald(2001)Bojowald]{mb} 
         Bojowald M., 2001, PhRvL, 86, 5227  
  
\bibitem[Borde et al.(2003)Borde, Guth, \& Vilenkin]{bgv} 
         Borde A., Guth A.~H., Vilenkin A., 2003, PhRvL, 90, 151301  
 
\bibitem[Bouhmadi-L{\'o}pez et al.(2008)]{mariam} 
         Bouhmadi-L{\'o}pez M., Gonz{\'a}lez-D{\'{\i}}az P.~F., Mart{\'{\i}}n-Moruno P., 2008, PhLB, 659, 1 
 
\bibitem[Cailleteau et al.(2008)Cailleteau et al.]{portsmouth} 
         Cailleteau T., Cardoso A., Vandersloot K., Wands D., 2008, PhRvL, 101, 251302 

\bibitem[Cailleteau et al.(2009)Cailleteau, Singh, \& Vandersloot]{csv} 
         Cailleteau T., Singh P., Vandersloot K., 2009, PhRvD, 80, 124013 
  
\bibitem[Caldwell et al.(2003)Caldwell, Kamionkowski, \& Weinberg]{caldwell} 
         Caldwell R.~R., Kamionkowski M., Weinberg N.~N., 2003, PhRvL, 91, 071301 

\bibitem[Chiou(2007)Chiou]{chiou} 
         Chiou D.-W., 2007, PhRvD, 75, 024029 

\bibitem[Chiou \& Vandersloot(2007)Chiou \& Vandersloot]{cv} 
         Chiou D.-W., Vandersloot K., 2007, PhRvD, 76, 084015 
 
\bibitem[Clarke \& Kr{\'o}lak(1985)Clarke \& Kr{\'o}lak]{clarke-krolak} 
         Clarke C.~J.~S., Kr{\'o}lak A., 1985, JGP, 2, 127  
 
\bibitem[Corichi \& Karami(2011)]{ck} Corichi A., Karami A., 2011, PhRvD, 83, 104006 
 
\bibitem[Corichi \& Karami(2014)]{ck-closed} Corichi A., Karami A., 2014, CQGra., 31, 035008 
  
\bibitem[Corichi \& Montoya(2011)Corichi \& Montoya]{cm1} Corichi A., Montoya E., 2011, PhRvD 84, 044021 
 
\bibitem[Corichi \& Singh(2008a)Corichi \& Singh]{cs1} Corichi A., Singh P., 2008a, PhRvL, 100, 161302   
 
\bibitem[Corichi \& Singh(2008b)Corichi \& Singh]{cs2} Corichi A., Singh P., 2008b, PhRvD, 78, 024034  
  
\bibitem[Corichi \& Singh(2009)]{cs3} Corichi A., Singh, P., 2009, PhRvD, 80, 044024   
  
\bibitem[Corichi \& Sloan (2013)]{corichisloan}  Corichi A., Sloan D, 2014, CQGra. 31, 062001
  
\bibitem[Craig \& Singh(2010)Craig \& Singh]{craig-singh1} Craig D.~A., Singh P., 2010, PhRvD, 82, 123526 
 
\bibitem[Craig \& Singh(2013)Craig \& Singh]{craig-singh2} Craig D.~A., Singh P., 2013, CQGra, 30, 205008 
 
\bibitem[de Risi, Maartens \& Singh (2007)]{pbb} de Risi G., Maartens R., Singh P., 2007, PhRvD, 76, 103531 
  
\bibitem[Diener et al.(2014a)Diener, Gupt, \& Singh]{dgs2} Diener P., Gupt B., Singh P., 2014, CQGra, 31, 105015   
  
\bibitem[Diener et al.(2014b)Diener et al.]{squeezed} Diener P., Gupt B., Megevand M., Singh P., 2014, CQGra, 31, 165006 

\bibitem[Diener et al.(2014c)Diener et al.]{cyclic} Diener P., Gupt B., Megevand M., Singh P., 2014, To appear 
 
\bibitem[Doroshkevich(1965)Doroshkevich]{dorosh} Doroshkevich A.~G., 1965, Ap, 1, 138 
 
\bibitem[Ellis(1967)Ellis]{ellis} Ellis G.~F.~R., 1967, JMP, 8, 1171 
  
\bibitem[Ellis \& MacCallum(1969)Ellis \& MacCallum]{ellis1} Ellis G.~F.~R., MacCallum M.~A.~H., 1969, CMaPh, 12, 108 
 
\bibitem[Ellis \& Schmidt(1977)Ellis \& Schmidt]{ellis-schmidt} Ellis G.~F.~R., Schmidt B.~G., 1977, GReGr, 8, 915 
 
\bibitem[Fern{\'a}ndez-Jambrina \& Lazkoz(2004)Fern{\'a}ndez-Jambrina \& Lazkoz]{lazkoz} 
         Fern{\'a}ndez-Jambrina L., Lazkoz R., 2004, PhRvD, 70, 121503 

\bibitem[Gambini \& Pullin (2011)Gambini \& Pullin]{pullin} Gambini R., Pullin J., 2011, 
         A first course in loop quantum gravity, Oxford University Press
 
\bibitem[Gambini \& Pullin(2008)]{gambini1} Gambini R., Pullin J., 2008, PhRvL, 101, 161301 
  
\bibitem[Gambini \& Pullin(2013)]{gambini2} Gambini R., Pullin J., 2013, PhRvL, 110, 211301 
  
\bibitem[Gambini \& Pullin(2014)]{gambini3} Gambini R., Pullin J., 2014, CQGra, 31, 115003 
  
\bibitem[Garay et al.(2010)Garay, Mart{\'{\i}}n-Benito, \& Mena Marug{\'a}n]{gowdy2} 
         Garay L.~J., Mart{\'{\i}}n-Benito M., Mena Marug{\'a}n G.~A., 2010, PhRvD, 82, 044048   
  
\bibitem[Garfinkle (2007)Garfinkle]{garfinkle} Garfinkle D., 2007, CQGra. 24, 295  

\bibitem[Garriga, Vilenin \& Zhang (2013)]{multiverse1} Garriga J., Vilenkin A., Zhang J., 2013, JCAP, 11, 55 

\bibitem[Gupt \& Singh(2012)]{gs1} Gupt B., Singh P., 2012, PhRvD, 85, 044011 

\bibitem[Gupt \& Singh(2013)]{gs2} Gupt B., Singh P., 2013, CQGra, 30, 145013 

\bibitem[Gupt \& Singh(2012)]{gs3} Gupt B., Singh P., 2012, PhRvD, 86, 024034 
  
\bibitem[Gupt \& Singh(2014)]{gs4} Gupt B., Singh P., 2014, PhRvD, 89, 063520 
  
\bibitem[Jacobs(1968)Jacobs]{jacobs} Jacobs K.~C., 1968, ApJ, 153, 661 
  
\bibitem[Joe \& Singh(2014)]{js1} Joe A., Singh P., 2014, arXiv:1407.2428  
 
\bibitem[Kaminski \& Pawlowski(2010a)Kaminski \& Pawlowski]{kp} Kaminski W., Pawlowski T., 2010a, PhRvD, 81, 024014  
 
\bibitem[Kaminski \& Pawlowski(2010b)Kaminski \& Pawlowski]{triangle} 
         Kaminski W., Pawlowski T., 2010b, PhRvD, 81, 084027 

\bibitem[Krolak(1986)Krolak]{krolak} Krolak A., 1986, CQGra, 3, 267   
 
\bibitem[Martin-Benito, Garay \& Marugan (2008)Martin-Benito, Garay \& Marugan]{gowdy1} 
         Martin-Benito M., Garay L. J., Marugan G. A. M., 2008, PhRvD 78, 083516  

\bibitem[Martin-Benito,  Marugan, Pawlowski (2009)Martin-Benito,  Marugan \& Pawlowski]{madrid_bianchi1} 
         Martin-Benito M., Mena Marugan G. A., Pawlowski T., 2009, PhRvD, 80, 084038

\bibitem[Naskar \& Ward(2007)]{naskar} Naskar T., Ward J., 2007, PhRvD 76, 063514 

\bibitem[Nojiri et al.(2005)Nojiri, Odintsov, \& Tsujikawa]{not} 
         Nojiri S., Odintsov S.~D., Tsujikawa S., 2005, PhRvD, 71, 063004 
  
\bibitem[Pawlowski, Ashtekar(2012)Pawlowski \& Ashtekar]{ap} 
         Pawlowski T., Ashtekar A., 2012, PhRvD, 85, 064001   
  
\bibitem[Pawlowski et al.(2014)Pawlowski, Pierini, \& Wilson-Ewing]{rad} 
         Pawlowski T., Pierini R., Wilson-Ewing E., 2014, arXiv:1404.4036   

\bibitem[Ranken \& Singh(2012)]{multi} Ranken E., Singh P., 2012, PhRvD 85, 104002 
  
\bibitem[Rovelli (2004)Rovelli]{rovelli} Rovelli C., 2004, Quantum Gravity, Cambridge University Press
  
\bibitem[Samart \& Gumjudpai(2007)Samart \& Gumjudpai]{gumju} 
         Samart D., Gumjudpai B., 2007, PhRvD, 76, 043514   
  
\bibitem[Sami et al.(2006)Sami, Singh, \& Tsujikawa]{sst} 
         Sami M., Singh P., Tsujikawa S., 2006, PhRvD, 74, 043514
  
\bibitem[Shtanov \& Sahni(2003)Shtanov \& Sahni]{ss} 
         Shtanov Y., Sahni V., 2003, PhLB, 557, 1   
  
\bibitem[Singh(2006)Singh]{ps06} Singh P., 2006, PhRvD, 73, 063508 
  
\bibitem[Singh(2009)Singh]{ps09} Singh P., 2009, CQGra, 26, 125005   
  
\bibitem[Singh(2012)Singh]{ps11} Singh P., 2012, PhRvD, 85, 104011   
  
\bibitem[Singh(2012)Singh]{ps12} Singh P., 2012, CQGra, 29, 244002   
  
\bibitem[Singh, Sami \& Dadhich(2003)Singh, Sami \& Dadhich]{ssd}  Singh P., Sami M., Dadhich N., 2003, Phys. Rev. D 68, 023522  
  
\bibitem[Singh, Vandersloot \& Vereshchagin(2006)Singh, Vandersloot \& Vereshchagin]{svv} 
         Singh P., Vandersloot K., Vereshchagin G.~V., 2006, PhRvD 74, 043510   
  
\bibitem[Singh \& Vidotto(2011)Singh \& Vidotto]{psfv} 
         Singh P., Vidotto F., 2011, PhRvD, 83, 064027 
  
\bibitem[Singh \& Wilson-Ewing(2014)Singh \& Wilson-Ewing]{pswe} 
         Singh P., Wilson-Ewing E., 2014, CQGra, 31, 035010   
 
\bibitem[Szulc et al.(2007) Szulc, Kaminski, \& Lewandowski]{warsaw} 
         Szulc {\L}., Kaminski W., Lewandowski J., 2007, CQGra, 24, 2621  
 
\bibitem[Szulc(2007)Szulc]{szulc} Szulc {\L}., 2007, CQGra, 24, 6191 

\bibitem[Tarr{\'{\i}}o et al.(2013)]{madridgowdy} 
         Tarr{\'{\i}}o P., Fern{\'a}ndez-M{\'e}ndez M., Mena Marug{\'a}n G.~A., 2013, PhRvD, 88, 084050 

\bibitem[Taveras(2008)Taveras]{vt} Taveras V., 2008, PhRvD, 78, 064072 
  
\bibitem[Thiemann(2007)]{thiemann} Thiemann, T.\ 2007, Modern Canonical Quantum General Relativity, 
         Cambridge University Press

\bibitem[Thorne(1967)Thorne]{thorne} Thorne K.~S., 1967, ApJ, 148, 51 
  
\bibitem[Tipler(1977)Tipler]{tipler} Tipler F.~J., 1977, PhLA, 64, 8   
  
\bibitem[Vandersloot(2007)Vandersloot]{open} Vandersloot K., 2007, PhRvD, 75, 023523   
  
\bibitem[Catto{\"e}n \& Visser(2005)Catto{\"e}n \& Visser]{visser} 
         Catto{\"e}n C., Visser M., 2005, CQGra, 22, 4913   
  
\bibitem[Willis(2004)Willis]{jw} Willis J.~L., 2004, PhD Thesis, Pennsylvania State University  

\bibitem[Wilson-Ewing(2010)Wilson-Ewing]{we} Wilson-Ewing E., 2010, PhRvD, 82, 043508   
  
\end{thebibliography}

\end{document}